\documentclass[preprint,proceedings]{rmaa}
\usepackage{paralist}
\usepackage{psfrag,color}

\SetYear{2004}
\SetConfTitle{Gravitational Collapse: From Massive Stars to Planets}

\title{Photon Bubbles in Young Massive Stars}

\author{
  N. J. Turner,\altaffilmark{1}
  H. W. Yorke,\altaffilmark{1}
  A. Socrates,\altaffilmark{2} and
  O. M. Blaes\altaffilmark{2}
}

\altaffiltext{1}{Jet Propulsion Laboratory, California Institute of
  Technology.}
\altaffiltext{2}{University of California at Santa Barbara.}

\shortauthor{Turner et al.}
\shorttitle{Photon Bubbles in Young Massive Stars}

\fulladdresses{
\item N. J. Turner and H. W. Yorke: MS 169-506, Jet Propulsion
Laboratory, California Institute of Technology, Pasadena CA 91109,
USA; \email{neal.turner@jpl.nasa.gov}.
\item O. M. Blaes and A. Socrates: Physics Department, University of
California, Santa Barbara CA 93106, USA.
}

\listofauthors{N. J. Turner, H. W. Yorke, A. Socrates \& O. M. Blaes}
\indexauthor{Turner, N. J.}
\indexauthor{Yorke, H. W.}
\indexauthor{Socrates, A.}
\indexauthor{Blaes, O. M.}

\abstract{Spectroscopic studies indicate that gas in the photospheres
of young O stars moves at speeds up to the sound speed.  We show,
using two-dimensional radiation MHD calculations and results from a
local linear analysis, that the motions may be due to photon bubble
instability if young O stars have magnetic fields.  }

\resumen{ }

\addkeyword{stars: early-type}
\addkeyword{stars: formation}

\def\msol{\rm\,M_\odot}
\def\divv{{\bf\nabla\cdot v}}

\begin{document}
\maketitle

\section{Introduction}

Young high-mass stars may trigger or terminate nearby star and planet
formation through ionizing radiation and winds launched from their
surfaces.  The properties of the radiation and winds depend on the
structure of the stellar surface layers.  Stars of more than 15 Solar
masses reach the main sequence while still accreting material
\citep{ys02}.  High-mass main sequence stars are thought to have
stably stratified outer layers, yet gas motions are present in the
atmospheres of these stars.

Absorption lines formed in the photospheres of O stars are broader
than expected based on the temperature, pressure and stellar rotation.
If the rotation axes are randomly oriented with respect to our line of
sight, some stars will be viewed pole-on and show no rotational
broadening.  However samples of O stars appear to contain no such
pole-on cases \citep{s56,ce77,p96,hs97}.  The minimum apparent
equatorial rotation speed is a function of spectral type and
luminosity.  Among main-sequence O stars in the Small Magellanic Cloud
cluster NGC 346, the excess broadening of UV metal lines ranges from
25 km s$^{-1}$ at O2, similar to the speed of sound at the
photosphere, to 5 km s$^{-1}$ at O9.5 \citep{bl03}.

Processes that might be responsible for the additional line
broadening include
\begin{enumerate}
\item {\em Vertical velocity gradients in the stellar wind
acceleration region} \citep{k92}: Minimum line widths are similar in
stars with and without strong outflows.  Also, stellar atmosphere plus
wind calculations including a variation in velocity through the
acceleration region produce photospheric lines similar to those in
hydrostatic calculations \citep{bl03}.
\item {\em Global non-radial pulsations:} Low levels of spectral line
shape variations in main-sequence O stars \citep{fg96} suggest that
any global non-radial modes are either of low amplitude or high order.
An unresolved issue is how such modes might be excited.
\item {\em Strange modes:} Instability occurs only for masses greater
than $80\msol$ in zero-age main sequence stars with metallicity
$Z=0.02$ \citep{gk93}, while excess photospheric line widths are
observed in main-sequence stars with masses down to $20\msol$.
\item {\em Shaviv modes:} The instabilities are found only at
luminosities greater than half the Eddington value \citep{s01}, while
excess line broadening occurs in stars with one-tenth the Eddington
luminosity.
\end{enumerate}

\begin{figure}[t!]
  \vspace*{-6mm} \hspace*{-3mm}
  \includegraphics[width=0.95\columnwidth]{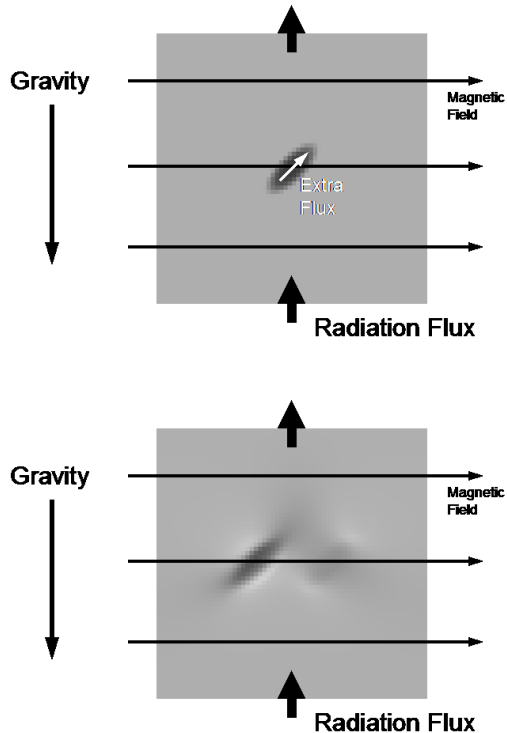} \vspace*{-1mm}
  \caption{How photon bubbles work.  {\em Top:} A patch of atmosphere
  (gray square) is initially in hydrostatic balance.  When density is
  reduced slightly in a small elongated region, photons diffuse more
  easily up the long axis.  The perturbed radiation flux exerts an
  extra force with a component along the magnetic field.  Gas is
  driven along the field to the right, out of the region of low
  density, and the perturbation grows.  {\em Bottom:} As fresh gas
  enters the region of low density from the left, the pattern
  propagates to the left at about the gas sound speed.  }
  \label{fig:howitworks}
\end{figure}

Here we explore the possibility that a fifth mechanism, photon bubble
instability \citep{a92,g98}, may lead to small-scale motions in the
surface layers of O stars.  A general local WKB analysis of linear
disturbances in optically-thick radiating atmospheres indicates
instability under a wide range of conditions \citep{bs03}.  Photon
bubbles are driven by the radiative flux
(Figure~\ref{fig:howitworks}), and grow if (1) Rosseland mean optical
depth per wavelength is greater than unity, (2) a magnetic field is
present, and (3) the radiative flux exceeds a critical value that in
radiation-supported atmospheres is approximately the gas sound speed
times the sum of gas and photon energy densities.  The surface layers
of O stars generally satisfy the first and third criteria, but until
recently were thought to be unmagnetized.  However $\theta^1$~Ori~C,
the illuminating star of the Orion Nebula and one of the nearest young
O stars, shows polarimetric variations in photospheric lines with
rotational phase, consistent with a 1.1~kG dipole field inclined
$42^\circ$ from the rotation axis \citep{db02}.  Also, narrow X-ray
emission lines indicate some gas near the star moves more slowly than
the wind.  The line ratios show that most of the X-ray flux is thermal
emission from plasma hotter than 15 million~K, too hot to arise in
shocks in the wind, and as hot as coronae in some magnetically active
lower-mass stars \citep{sc03}.  Overall, there is good evidence for a
magnetic field in $\theta^1$~Ori~C.

Photon bubbles, unlike strange modes, can grow in scattering
atmospheres.  They differ from Shaviv instabilities in that they
require magnetic fields and may grow when the luminosity is
substantially less than the Eddington value.

\section{Methods}

A small patch of the surface layers of the star is followed in
Cartesian coordinates $(x, y, z)$.  The $x$-axis is horizontal, the
$z$-axis vertical, and symmetry is assumed along the $y$-axis.  The
domain is $10^{10}$~cm on a side, or about 1.5\% of the radius of
$\theta^1$ Ori C, and divided into $32\times 32$ zones.  The top of
the domain is placed at the photosphere.  The side boundaries are
periodic, and the upper and lower boundaries are transparent walls
through which radiation may pass, but gas and magnetic fields do not.
The temperature of the lower boundary is fixed at its initial value,
while the radiative flux is continuous across the upper boundary.  The
equations of radiation MHD,
\begin{equation}\label{eqn:cty}
{D\rho\over D t}+\rho\divv=0,
\end{equation}
\begin{equation}\label{eqn:eomg}
\rho{D{\bf v}\over D t} = -{\bf\nabla}p
	+ {1\over 4\pi}({\bf\nabla\times B}){\bf\times B}
        + {\chi_F\rho\over c}{\bf F} - \rho g {\bf\hat z},
\end{equation}
\begin{equation}\label{eqn:eoeg}
\rho{D\over D t}\left({e\over\rho}\right) =
	- p\divv - \kappa_P\rho(4\pi B - c E),
\end{equation}
\begin{equation}\label{eqn:eoer}
\rho{D\over D t}\left({E\over\rho}\right) =
	- {\bf\nabla\cdot F} - {\bf\nabla v}:\mathrm{P}
	+ \kappa_P\rho(4\pi B - c E),
\end{equation}
and
\begin{equation}\label{eqn:dbdt}
{\partial{\bf B}\over\partial t} = {\bf\nabla\times}({\bf v\times B}),
\end{equation}
\citep{mm84,sm92,bs03} are solved using the ZEUS code
\citep{sn92a,sn92b} with its flux-limited radiation diffusion module
\citep{ts01}.  A uniform gravity $g$ is included.  Rotation is
ignored, as photon bubbles grow to saturation within a typical stellar
rotation period.  Emission proportional to the blackbody rate $B$ is
assumed in equations~\ref{eqn:eoeg} and~\ref{eqn:eoer}.  Temperatures
are calculated using a mean mass per particle of $0.6$ protons.  The
total flux-mean opacity $\chi_F$ includes contributions from electron
scattering 0.34 cm$^2$ g$^{-1}$ and bound-free and free-free processes
$2\times 10^{52} \rho^{4.5}e^{-3.5}$ cm$^2$ g$^{-1}$, where $e$ is the
gas internal energy per unit volume and $\rho$ is the mass density.
The Planck-mean absorption opacity $\kappa_P$ due to the bound-free
and free-free effects is $7.4\times 10^{53} \rho^{4.5}e^{-3.5}$ cm$^2$
g$^{-1}$.  The radiative flux ${\bf F} =
-{c\Lambda\over\chi_F\rho}{\bf\nabla}E$ is calculated using a limiter
$\Lambda$ that equals $\frac{1}{3}$ in optically-thick regions, and is
reduced in regions of low optical depth so that the flux does not
exceed the product of the radiation energy density $E$ and light speed
$c$ \citep{lp81}.  The set of equations is closed using an ideal gas
equation of state $p=(\gamma-1)e$ with $\gamma=5/3$.

Initial conditions are constructed by integrating the equations of
hydrostatic equilibrium and radiative flux conservation from the
photosphere down into the interior.  Three initial states are
considered, spanning the range of effective temperatures found in O
stars.  The photospheric temperatures in Kelvins and densities in g
cm$^{-3}$ are (A) $50\,000, 1.8\times 10^{-9}$, (B) $40\,000,
2.2\times 10^{-9}$, and (C) $30\,000, 3.0\times 10^{-9}$.  Surface
gravity $10^{3.8}$ cm s$^{-2}$ is used in all cases.  The Eddington
ratios between the last two terms in equation~\ref{eqn:eomg} at the
photospheres are (A) $0.65$, (B) $0.28$, and (C) $0.11$.  We study the
growth of photon bubbles at the stellar magnetic equator by beginning
with a uniform, horizontal magnetic field of 1~kG.  Radiation, gas and
magnetic pressures at the photosphere are in ratios (A) $0.40:0.31:1$,
(B) $0.16:0.31:1$, and (C) $0.05:0.31:1$.  Random zone-to-zone initial
density perturbations with amplitude 0.1\% are added only in the
middle half of the domain height, to reduce any effects of the
boundaries on the linear growth.

\section{Results}

Each calculation passes through three stages.  First, the
perturbations cause small damped vertical oscillations with period
equal to the gas sound crossing time and amplitude 0.1 km s$^{-1}$.
Next, linear eigenmodes appear and grow exponentially.  The
fastest-growing modes are gas sound waves with wavefronts inclined 30
to 45 degrees from vertical, and group velocity roughly parallel to
the magnetic field.  Finally, the growth saturates and amplitudes vary
little over many gas sound crossing times.

\subsection{Linear Growth}

\begin{figure}[t!]
  \hspace*{1mm} \includegraphics[width=0.91\columnwidth]{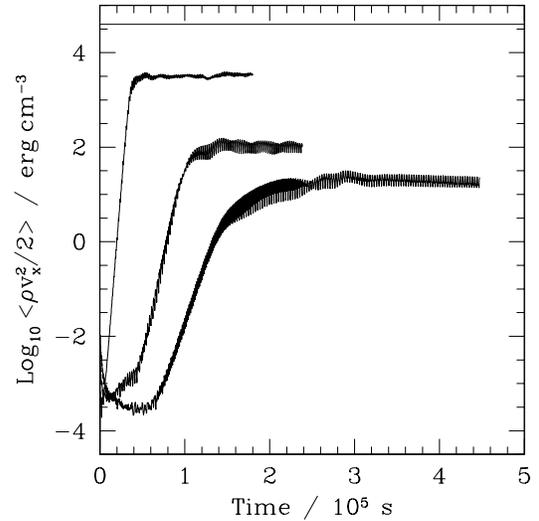}
  \caption{Domain-averaged kinetic energies in horizontal motions
  versus time, in the calculations with photospheric temperatures
  $50\,000$ (top), $40\,000$ (middle), and $30\,000$~K (bottom).  The
  upper horizontal line indicates the energy density in the initial
  magnetic field.  } \label{fig:time}
\end{figure}

The time histories of the horizontal kinetic energies are shown in
Figure~\ref{fig:time}.  Linear growth rates increase with photospheric
temperature.  In the $50\,000$~K case (A), the fastest mode has 3
wavelengths in the domain width and 2 in the height, corresponding to
wavelength $2.8\times 10^9$ cm and wavevector $34^\circ$ from
horizontal.  The power in the mode, measured using the Fourier
transform of the horizontal velocity pattern, grows $e$-fold every
3000~seconds.  The growth rate is $2/3$ that obtained by solving the
linear dispersion relation, Blaes \& Socrates (2003) equation~49, for
conditions at the center of the computational domain.  The slower
growth in the simulation may be due to the modest grid resolution.

\subsection{Saturation}

\begin{figure}[t!]
  \hspace*{12mm} \includegraphics[width=0.67\columnwidth]{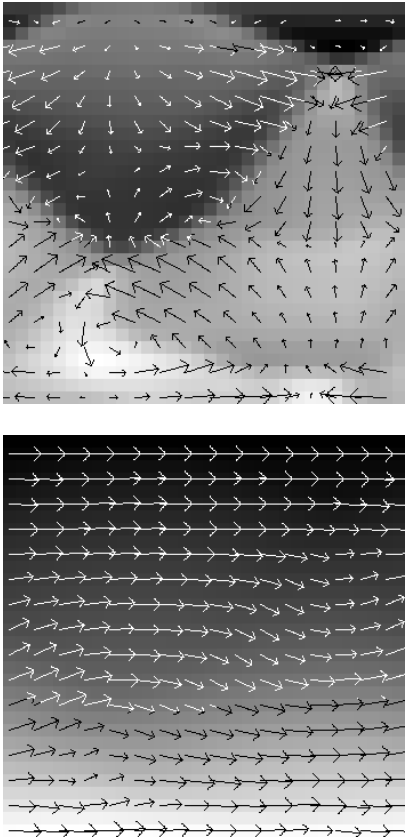}
  \caption{Snapshot of saturated photon bubble instability in the
  calculation with photospheric temperature $50\,000$~K, at
  $56\,000$~seconds.  The initial photosphere lies along the top
  boundary, and the domain height and width are $10^{10}$~cm.  {\em
  Top:} Density on a gray scale logarithmic between $1.2\times
  10^{-9}$ (black) and $1.3\times 10^{-8}$ g cm$^{-3}$ (white).
  Velocities are shown by arrows, the longest corresponding to 29 km
  s$^{-1}$.  {\em Bottom:} Radiation pressure on a gray scale linear
  between $6\,600$ and $210\,000$ dyn cm$^{-2}$.  Magnetic fields are
  shown by arrows, the longest 1500~G.  } \label{fig:saturated}
\end{figure}

The disturbances reach larger amplitudes in the cases with hotter
photospheres.  The largest horizontal velocities are $30$ km~s$^{-1}$
in the $50\,000$~K case (A), $12$ km~s$^{-1}$ in the $40\,000$~K case
(B), and $6$ km~s$^{-1}$ in the $30\,000$~K case (C).  In the hottest
case (A), the speeds are similar to the isothermal gas sound speed of
$26$ km~s$^{-1}$ at the photosphere, and weak shocks are present.  The
structure resembles that proposed by Begelman (2001), but with crossed
shocks due to the initial horizontal symmetry.  Gas oscillates back
and forth along magnetic field lines, experiencing repeated
compression and expansion.  Densities vary about their initial values
by up to a factor three, while radiation energy remains almost uniform
due to rapid diffusion (Figure~\ref{fig:saturated}).  Radiation
escapes more easily through the inhomogeneous atmosphere than through
the original hydrostatic structure, and the time-averaged radiative
flux is 10\% greater than initially.  The density variations in the
$40\,000$~K case (B) are 50\%, and in the $30\,000$~K case (C) are
20\%.  The magnetic fields bend $30^\circ$ up and down in the hottest
case, and remain close to horizontal in the two cooler cases.

\section{Conclusions}

Photon bubble instability is present in the surface layers of
magnetized main-sequence O stars according to the results of a local
linear WKB analysis.  We use two-dimensional radiation MHD
calculations with horizontal magnetic fields and closed vertical
boundaries to show that the instability can lead to small-scale
movements of gas in the surface layers, with velocities about equal to
the observed linewidth excesses.  In a calculation with parameters
appropriate for an early O star, photon bubbles result in horizontal
density variations near the photosphere.  Radiation diffuses more
rapidly through the inhomogeneities than through the hydrostatic
atmosphere with the same column depth, and the radiative flux is 10\%
greater than that calculated assuming hydrostatic equilibrium.  An
enhanced flux
may affect measurements of stellar parameters.  Issues remaining to be
resolved
\linebreak \adjustfinalcols
include the dependence on the local magnetic field
orientation and the shapes of the spectral lines expected from
emission integrated over the stellar disk.

Photon bubbles reaching amplitudes similar to those in the hottest
case we considered may eject blobs of gas through the photosphere, and
could lead to density variations in the winds of O stars.  The motions
might transfer energy into an initially weak magnetic field.
Additional radiation MHD calculations indicate the saturation
amplitude generally increases with decreasing surface gravity, so
photon bubbles may reach larger amplitudes in O giants and
supergiants.  Solutions of the dispersion relation show that photon
bubbles may be present also in accretion disks around young massive
stars.

\acknowledgements

N. J. T. was supported by a National Research Council fellowship.  The
work was carried out in part at the Jet Propulsion Laboratory,
California Institute of Technology.

\end{document}